\documentstyle[aps,twocolumn]{revtex}
\begin{document}

%
%
%
\edef\catcodeat{\the\catcode`\@ }     \catcode`\@=11
\newbox\p@ctbox                       
\newbox\t@mpbox                       
\newbox\@uxbox                        
\newbox\s@vebox                       
\newtoks\desct@ks \desct@ks={}        
\newtoks\@ppenddesc                   
\newtoks\sh@petoks                    
\newif\ifallfr@med  \allfr@medfalse   
\newif\if@ddedrows                    
\newif\iffirstp@ss  \firstp@ssfalse   
\newif\if@mbeeded                     
\newif\ifpr@cisebox                   
\newif\ifvt@p                         
\newif\ifvb@t                         
\newif\iff@nished    \f@nishedtrue    
\newif\iffr@med                       
\newif\ifj@stbox     \j@stboxfalse    
\newcount\helpc@unt                   
\newcount\p@ctpos                     
\newdimen\r@leth      \r@leth=0.4pt   
\newdimen\x@nit                       
\newdimen\y@nit                       
\newdimen\xsh@ft                      
\newdimen\ysh@ft                      
\newdimen\@uxdimen                    
\newdimen\t@mpdimen                   
\newdimen\t@mpdimeni                  
\newdimen\b@tweentandp                
\newdimen\b@ttomedge                  
\newdimen\@pperedge                   
\newdimen\@therside                   
\newdimen\d@scmargin                  
\newdimen\p@ctht                      
\newdimen\l@stdepth
\newdimen\in@tdimen
\newdimen\l@nelength
\newdimen\re@lpictwidth
\def\justframes{\global\j@stboxtrue}  
\def\picturemargins#1#2{\b@tweentandp=#1\@therside=#2\relax}
\def\allframed{\global\allfr@medtrue} 
\def\emptyplace#1#2{\pl@cedefs        
    \setbox\@uxbox=\vbox to#2{\n@llpar
        \hsize=#1\vfil \vrule height0pt width\hsize}
    \e@tmarks}
\def\boxplace{\pl@cedefs\afterassignment\re@dvbox\let\n@xt= }
\def\re@dvbox{\setbox\@uxbox=\vbox\bgroup
         \n@llpar\aftergroup\e@tmarks}
\def\fontcharplace#1#2{\pl@cedefs     
    \setbox\@uxbox=\hbox{#1\char#2\/}%
    \xsh@ft=-\wd\@uxbox               
    \setbox\@uxbox=\hbox{#1\char#2}%
    \advance\xsh@ft by \wd\@uxbox     
    \helpc@unt=#2
    \advance\helpc@unt by -63         
    \x@nit=\fontdimen\helpc@unt#1%
    \advance\helpc@unt by  20         
    \y@nit=\fontdimen\helpc@unt#1%
    \advance\helpc@unt by  20         
    \ifnum\helpc@unt<51
      \ysh@ft=-\fontdimen\helpc@unt#1%
    \fi
    \e@tmarks}
\def\n@llpar{\parskip0pt \parindent0pt
    \leftskip=0pt \rightskip=0pt
    \everypar={}}
\def\pl@cedefs{\xsh@ft=0pt\ysh@ft=0pt}
\def\e@tmarks#1{\setbox\@uxbox=\vbox{ 
      \n@llpar
      \hsize=\wd\@uxbox               
      \noindent\copy\@uxbox           
      \kern-\wd\@uxbox                
      #1\par}
    \st@redescription}
\def\t@stprevpict#1{\ifvoid#1\else    
   \errmessage{Previous picture is not finished yet.}\fi} 

\def\st@redescription#1\par{
    \global\setbox\s@vebox=\vbox{\box\@uxbox\unvbox\s@vebox}%
    \desct@ks=\expandafter{\the\desct@ks#1\@ndtoks}}
\def\def@ultdefs{\p@ctpos=1         
      \def\lines@bove{0}
      \@ddedrowsfalse               
      \@mbeededfalse                
      \pr@ciseboxfalse
      \vt@pfalse                    
      \vb@tfalse                    
      \@ppenddesc={}
      \ifallfr@med\fr@medtrue\else\fr@medfalse\fi
      }

\def\descriptionmargins#1{\global\d@scmargin=#1\relax}
\def\@dddimen#1#2{\t@mpdimen=#1\advance\t@mpdimen by#2#1=\t@mpdimen}
\def\placemark#1#2 #3 #4 #5 {\unskip    
      \setbox1=\hbox{\kern\d@scmargin#5\kern\d@scmargin}
      \@dddimen{\ht1}\d@scmargin        
      \@dddimen{\dp1}\d@scmargin        
      \ifx#1l\dimen3=0pt\else           
        \ifx#1c\dimen3=-0.5\wd1\else
          \ifx#1r\dimen3=-\wd1
     \fi\fi\fi
     \ifx#2u\dimen4=-\ht1\else          
       \ifx#2c\dimen4=-0.5\ht1\advance\dimen4 by 0.5\dp1\else
         \ifx#2b\dimen4=0pt\else
           \ifx#2l\dimen4=\dp1
     \fi\fi\fi\fi
     \advance\dimen3 by #3
     \advance\dimen4 by #4
     \advance\dimen4 by-\dp1
     \advance\dimen3 by \xsh@ft         
     \advance\dimen4 by \ysh@ft         
     \kern\dimen3\vbox to 0pt{\vss\copy1\kern\dimen4}
     \kern-\wd1                        
     \kern-\dimen3                     
     \ignorespaces}                    
\def\fontmark #1#2 #3 #4 #5 {\placemark #1#2 #3\x@nit{} #4\y@nit{} {#5} }
\def\fr@msavetopict{\global\setbox\s@vebox=\vbox{\unvbox\s@vebox
      \global\setbox\p@ctbox=\lastbox}%
    \expandafter\firstt@ks\the\desct@ks\st@ptoks}
\def\firstt@ks#1\@ndtoks#2\st@ptoks{%
    \global\desct@ks={#2}%
    \def\t@mpdef{#1}%
    \@ppenddesc=\expandafter\expandafter\expandafter
                        {\expandafter\t@mpdef\the\@ppenddesc}}
\def\testf@nished{{\let\s@tparshape=\relax
    \s@thangindent}}
\def\inspicture{\t@stprevpict\p@ctbox
    \def@ultdefs                  
    \fr@msavetopict
    \iff@nished\else\testf@nished\fi
    \iff@nished\else
      \immediate\write16{Previes picture is not finished yet}%
    \fi
    \futurelet\N@xt\t@stoptions}  
\def\t@stoptions{\let\n@xt\@neletter
  \ifx\N@xt l\p@ctpos=0\else                
   \ifx\N@xt c\p@ctpos=1\else               
    \ifx\N@xt r\p@ctpos=2\else              
     \ifx\N@xt(\let\n@xt\e@tline\else        
      \ifx\N@xt!\@mbeededtrue\else           
       \ifx\N@xt|\fr@medtrue\else            
        \ifx\N@xt^\vt@ptrue\vb@tfalse\else  
         \ifx\N@xt_\vb@ttrue\vt@pfalse\else 
          \ifx\N@xt\bgroup\let\n@xt\@ddgrouptodesc\else
           \let\n@xt\@dddescription 
  \fi\fi\fi\fi\fi\fi\fi\fi\fi\n@xt}
\def\e@tline(#1){\def\lines@bove{#1}
    \@ddedrowstrue
    \futurelet\N@xt\t@stoptions}
\def\@neletter#1{\futurelet\N@xt\t@stoptions} 
\def\@ddgrouptodesc#1{\@ppenddesc={#1}\futurelet\N@xt\t@stoptions}
\def\fr@medpict{\setbox\p@ctbox=
    \vbox{\n@llpar\hsize=\wd\p@ctbox
       \iffr@med\else\r@leth=0pt\fi
       \ifj@stbox\r@leth=0.4pt\fi
       \hrule height\r@leth \kern-\r@leth
       \vrule height\ht\p@ctbox depth\dp\p@ctbox width\r@leth \kern-\r@leth
       \ifj@stbox\hfill\else\copy\p@ctbox\fi
       \kern-\r@leth\vrule width\r@leth\par
       \kern-\r@leth \hrule height\r@leth}}
\def\@dddescription{\fr@medpict     
    \re@lpictwidth=\the\wd\p@ctbox
    \advance\re@lpictwidth by\@therside
    \advance\re@lpictwidth by\b@tweentandp
    \ifhmode\ifinner\pr@ciseboxtrue\fi\fi
    \createp@ctbox
    \let\N@xt\tr@toplacepicture
    \ifhmode                         
      \ifinner\let\N@xt\justc@py
      \else\let\N@xt\vjustc@py
      \fi
    \else
      \ifnum\p@ctpos=1               
        \let\N@xt\justc@py
      \fi
    \fi
    \if@mbeeded\let\N@xt\justc@py\fi 
    \firstp@sstrue
    \N@xt}
\def\createp@ctbox{\global\p@ctht=\ht\p@ctbox
    \advance\p@ctht by\dp\p@ctbox
    \advance\p@ctht by 6pt
    \setbox\p@ctbox=\vbox{
      \n@llpar                     
      \t@mpdimen=\@therside          
      \t@mpdimeni=\hsize             
      \advance\t@mpdimeni by -\@therside
      \advance\t@mpdimeni by -\wd\p@ctbox
      \ifpr@cisebox
        \hsize=\wd\p@ctbox
      \else
        \ifcase\p@ctpos
               \leftskip=\t@mpdimen    \rightskip=\t@mpdimeni
        \or    \advance\t@mpdimeni by \@therside
               \leftskip=0.5\t@mpdimeni \rightskip=\leftskip
        \or    \leftskip=\t@mpdimeni   \rightskip=\t@mpdimen
        \fi
      \fi
      \hrule height0pt             
      \kern6pt                     
      \penalty10000
      \noindent\copy\p@ctbox\par     
      \kern3pt                       
      \hrule height0pt
      \hbox{}%
      \penalty10000
      \interlinepenalty=10000
      \the\@ppenddesc\par            
      \penalty10000                  
      \kern3pt                       
      }%
      \ifvt@p
       \setbox\p@ctbox=\vtop{\unvbox\p@ctbox}%
      \else
        \ifvb@t\else
          \@uxdimen=\ht\p@ctbox
          \advance\@uxdimen by -\p@ctht
          {\vfuzz=\maxdimen
           \global\setbox\p@ctbox=\vbox to\p@ctht{\unvbox\p@ctbox}%
          }%
          \dp\p@ctbox=\@uxdimen
        \fi
      \fi
      }
\def\picname#1{\unskip\setbox\@uxbox=\hbox{\bf\ignorespaces#1\unskip\ }%
      \hangindent\wd\@uxbox\hangafter1\noindent\box\@uxbox\ignorespaces}
\def\justc@py{\ifinner\box\p@ctbox\else\kern\parskip\unvbox\p@ctbox\fi
  \global\setbox\p@ctbox=\box\voidb@x}
\def\vjustc@py{\vadjust{\kern0.5\baselineskip\unvbox\p@ctbox}%
      \global\setbox\p@ctbox=\box\voidb@x}
\def\tr@toplacepicture{
      \ifvmode\l@stdepth=\prevdepth  
      \else   \l@stdepth=0pt         
      \fi
      \vrule height.85em width0pt\par
      \r@memberdims                  
      \global\t@mpdimen=\pagetotal
      \t@stheightofpage              
      \ifdim\b@ttomedge<\pagegoal    
         \let\N@xt\f@gurehere        
         \global\everypar{}
      \else
         \let\N@xt\relax             
         \penalty10000
         \vskip-\baselineskip        
         \vskip-\parskip             
         \immediate\write16{Picture will be shifted down.}%
         \global\everypar{\sw@tchingpass}
      \fi
      \penalty10000
      \N@xt}
\def\sw@tchingpass{
    \iffirstp@ss                     
      \let\n@xt\relax
      \firstp@ssfalse                
    \else
      \let\n@xt\tr@toplacepicture
      \firstp@sstrue
    \fi  \n@xt}
\def\r@memberdims{\global\in@tdimen=0pt
    \ifnum\p@ctpos=0
        \global\in@tdimen=\re@lpictwidth
      \fi
      \global\l@nelength=\hsize
      \global\advance\l@nelength by-\re@lpictwidth
      }
\def\t@stheightofpage{%
     \global\@pperedge=\t@mpdimen
     \advance\t@mpdimen by-0.7\baselineskip 
     \advance\t@mpdimen by \lines@bove\baselineskip 
     \advance\t@mpdimen by \ht\p@ctbox      
     \advance\t@mpdimen by \dp\p@ctbox      
     \advance\t@mpdimen by-0.3\baselineskip 
     \global\b@ttomedge=\t@mpdimen          
     }
\def\f@gurehere{\global\f@nishedfalse
      \t@mpdimen=\lines@bove\baselineskip   
      \advance\t@mpdimen-0.7\baselineskip   
      \kern\t@mpdimen
      \advance\t@mpdimen by\ht\p@ctbox
      \advance\t@mpdimen by\dp\p@ctbox
      {\t@mpdimeni=\baselineskip
       \offinterlineskip
       \unvbox\p@ctbox
       \global\setbox\p@ctbox=\box\voidb@x
       \penalty10000   \kern-\t@mpdimen     
       \penalty10000   \vskip-\parskip      
       \kern-\t@mpdimeni                    
      }%
      \penalty10000                         
      \global\everypar{\s@thangindent}
      }
\def\s@thangindent{%
    \ifdim\pagetotal>\b@ttomedge\global\everypar{}%
      \global\f@nishedtrue             
      \else
        \advance\@pperedge by -1.2\baselineskip
        \ifdim\@pperedge>\pagetotal\global\everypar{}%
          \global\f@nishedtrue
        \else
          \s@tparshape                 
        \fi
        \advance\@pperedge by 1.2\baselineskip
      \fi}
\def\s@tparshape{\t@mpdimen=-\pagetotal
   \advance\t@mpdimen by\b@ttomedge    
   \divide\t@mpdimen by\baselineskip   
   \helpc@unt=\t@mpdimen               
   \advance \helpc@unt by 2            
   \sh@petoks=\expandafter{\the\helpc@unt\space}
   \t@mpdimeni=\lines@bove\baselineskip
   \t@mpdimen=\pagetotal
   \gdef\lines@bove{0}
   \loop \ifdim\t@mpdimeni>0.999\baselineskip 
     \advance\t@mpdimen  by \baselineskip
     \advance\t@mpdimeni by-\baselineskip
     \sh@petoks=\expandafter{\the\sh@petoks 0pt \the\hsize}%
   \repeat
   \loop \ifdim\b@ttomedge>\t@mpdimen         
     \advance\t@mpdimen by \baselineskip
     \sh@petoks=\expandafter{\the\sh@petoks \in@tdimen \l@nelength }%
   \repeat
   \sh@petoks=\expandafter
      {\the\sh@petoks 0pt \the\hsize}
   \expandafter\parshape\the\sh@petoks
   }

\descriptionmargins{2pt}
\picturemargins{15pt}{0pt}

\catcode`\@=\catcodeat        \let\catcodeat=\undefined

\emptyplace{3.3in
\includegraphics{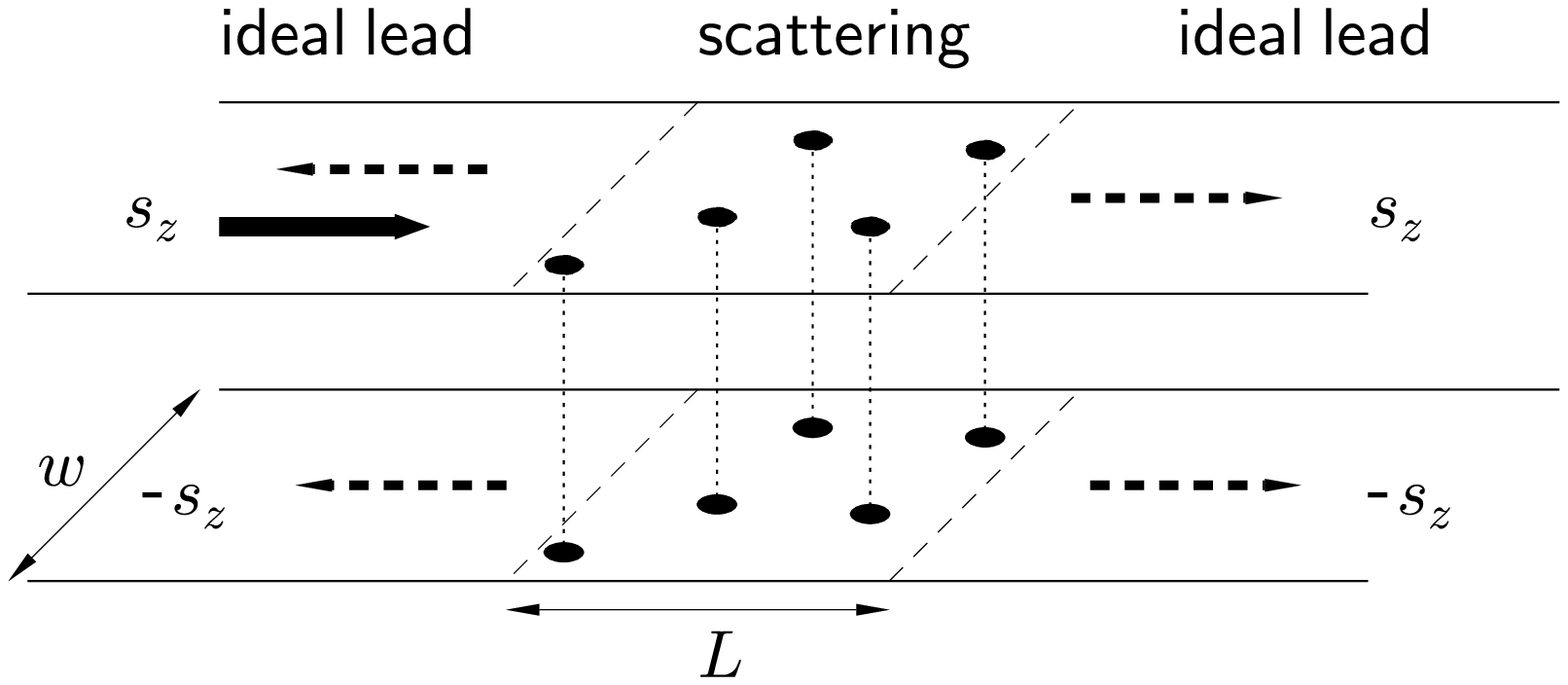}}
{1.7in}
{\footnotesize \noindent
FIG.1. Scheme of the scattering process in a two-component
system. Upper and lower strips of the width $w$ represent the
spin-subsystems. Scatterers, black points, serve also as connection
points between subsystems giving rise to a spin-flip processes.
Thick full and dashed lines represent an incoming and
outgoing waves, respectively.}

\emptyplace{3.3in
\includegraphics{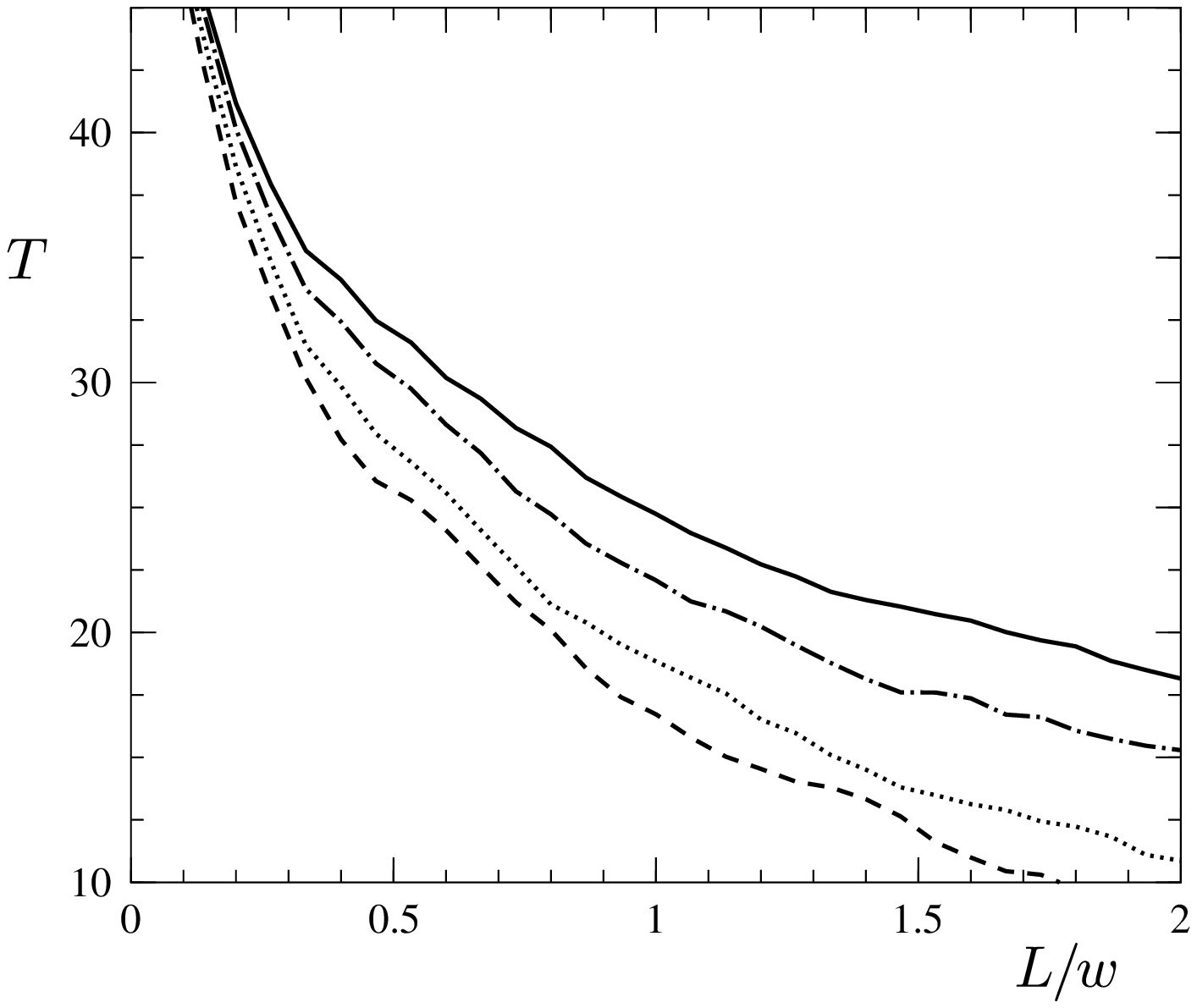}}
{2.8in}
{\footnotesize \noindent
FIG.2. The total transmission coefficient $T$ as function of the
scattering region length $L$ for several values of the
spin-flip probability:
$t^{\uparrow \downarrow} = 8.1 \times 10^{-3}$ (full line),
$t^{\uparrow \downarrow} = 4.2 \times 10^{-3}$ (dashed-dotted line),
$t^{\uparrow \downarrow} = 1.1 \times 10^{-3}$ (dotted line),
$t^{\uparrow \downarrow} = 0.0005 \times 10^{-3}$ (dashed line)
.}

\emptyplace{3.3in
\includegraphics{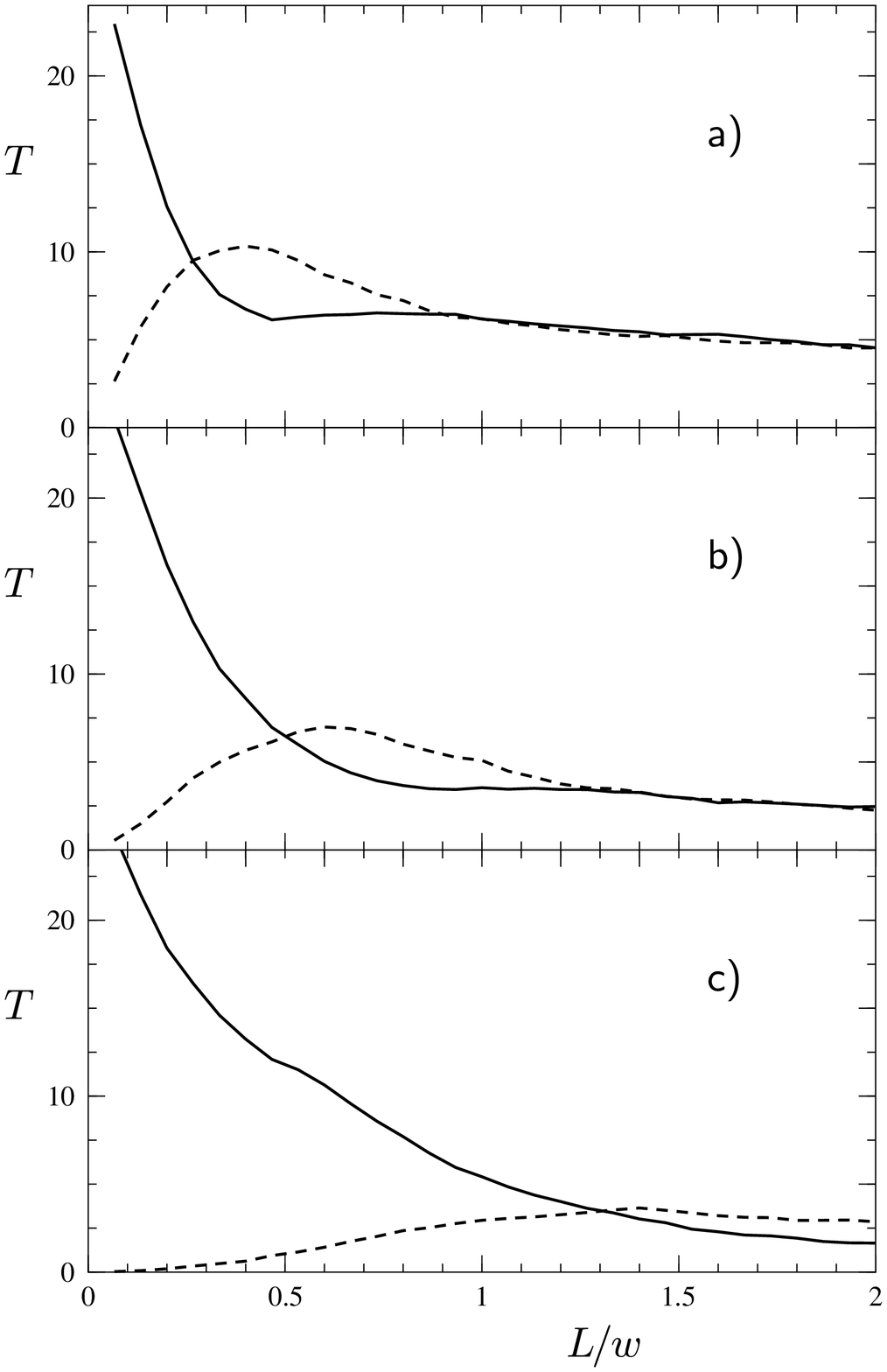}}
{5.0in}
{\footnotesize \noindent
FIG.3. The partial transmission coefficients $T^{\uparrow \uparrow}$
(full line) and $T^{\uparrow \downarrow}$ (dashed line) as function of
the scattering region length $L$ for several values of the
spin-flip probability:
a) $t^{\uparrow \downarrow} = 8.1 \times 10^{-3}$;
b) $t^{\uparrow \downarrow} = 0.2 \times 10^{-3}$;
c) $t^{\uparrow \downarrow} = 0.0005 \times 10^{-3}$.}

\emptyplace{3.3in
\includegraphics{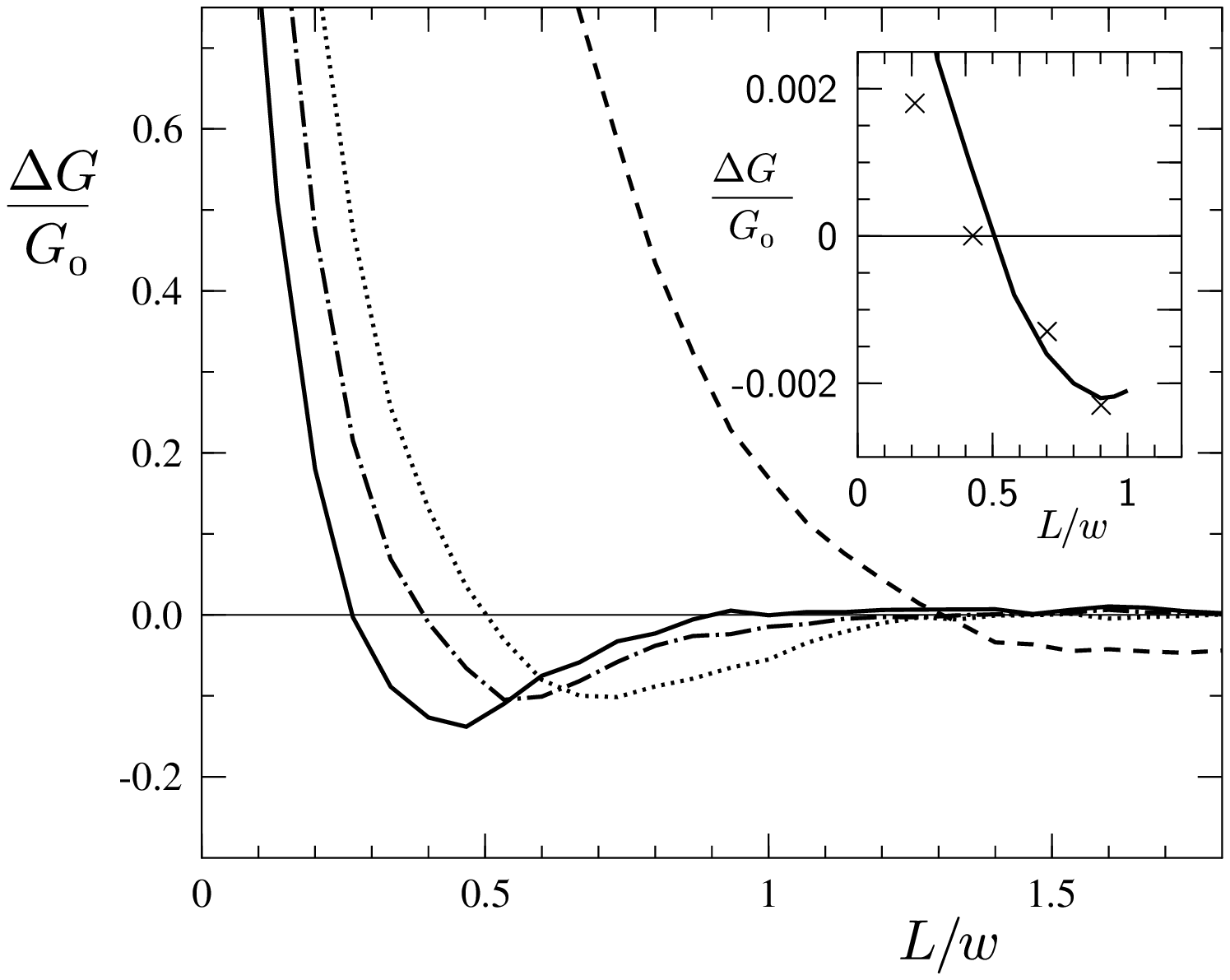}}
{2.8in}
{\footnotesize \noindent
FIG.4. Relative conductance change $\Delta G / G_{0}$
as function of the scattering region length $L$ for several
values of the spin-flip probability:
$t^{\uparrow \downarrow} = 8.1 \times 10^{-3}$  (full line),
$t^{\uparrow \downarrow} = 1.1 \times 10^{-3}$  (dashed-dotted line),
$t^{\uparrow \downarrow} = 0.2 \times 10^{-3}$  (dotted line),
$t^{\uparrow \downarrow} = 0.0005 \times 10^{-3}$  (dashed line).
In the inset crosses represent experimental data obtained by Hu
et al. and full line is the result of the model calculation for
the following parameters: $N=173$, $\alpha_s=0.01$,
$\beta_s=0$, scatterer concentration $1500/w^2$,
$\sigma_0=0.1341$ and $t^{\uparrow \downarrow}=4.8 \times 10^{-3}$.}

\title{Two-component model of a spin-polarized transport}

\author{P. \v{S}eba$^{1,3,4}$, P. Exner$^{2,3}$, K. N. Pichugin$^{1}$
and P. St\v{r}eda$^{1}$}
\address{$^{1}$Institute of Physics, Academy of Sciences of the
Czech Republic, Cukrovarnick\'{a} 10, 162 53 Praha \\
$^{2}$Nuclear Physics Institute, Academy of Sciences of the
Czech Republic, 250 68 \v{R}e\v{z} u Prahy \\
$^{3}$Doppler Institute, Czech Technical University,
B\v{r}ehov\'{a} 7, 115 19 Praha 1 \\
$^{4}$Department of Physics, Pedagogical University, V\'{\i}ta
Nejedl\'{e}ho 573, Hradec Kr\'{a}lov\'{e}}

\maketitle

\date{\today}

\begin{abstract}
Effect of the spin-involved interaction of electrons with impurity atoms
or defects to the transport properties of a two-dimensional
electron gas is described by using a simplifying two-component model.
Components representing spin-up and spin-down states are supposed
to be coupled at a discrete set of points
within a conducting channel. The used limit of the short-range
interaction allows to solve the relevant scattering problem exactly.
By varying the model parameters different transport regimes of
two-terminal devices with ferromagnetic contacts can be described.
In a quasi-ballistic regime the resulting
difference between conductances for the parallel and antiparallel
orientation of the contact magnetization changes its sign as a
function of the length of the conducting channel if appropriate
model parameters are chosen. This effect is in agreement
with recent experimental observation.
\end{abstract}

\narrowtext


\vspace*{-5mm}
Spin-polarized transport in two-dimensional electron systems
has been a field of growing interest during the last years.
Typically, the experiments are performed using a two-terminal
device with ferromagnetic metal contacts. A spin-polarization of
the injected current is expected from the different densities of
states for spin-up and spin-down electrons in the ferromagnetic
source. This leads to a spin dependent interface-resistance,
which also exists at the interface of the second ferromagnetic
contact, the drain. Together with spin-involved scattering
processes in the studied electron system this should result
in a conductance which
depends on the relative magnetization of the two contacts
\cite{datta}.

The quantum mechanical nature of spin places it out of reach of
many of the forces in a solid and the orientation of a carrier's
spin can be very long-lived. The conductance $G^{\uparrow \uparrow}$
of a two-terminal device with parallel orientation of magnetic
moments of the contacts is thus expected to be higher than the
conductance $G^{\uparrow \downarrow}$ for the case of antiparallel
moment orientation \cite{datta,johnson}. However, just the opposite
results have been reported recently \cite{hu} for a two-dimensional
electron gas confined in an InAs channel with the permalloy source and
drain. It has been found that an ensemble average of the conductance
difference $G^{\uparrow \uparrow} - G^{\uparrow \downarrow}$
decreases  as function of the channel length reaching negative
values in a quasi-ballistic regime when the electron mean free
path $l_e$ becomes comparable with the channel length.

In the absence of magnetic impurities the natural candidate for
spin dephasing and precession effects is spin-orbit coupling to
impurity atoms or defects. General theoretical approach to its
description is to use the contribution to the Hamiltonian
which stems directly from the quadratic in $v/c$ expansion of the
Dirac equation \cite{davydov}
\begin{equation}
\label{soham}
\hat{H}_{SO} \, = \,- \frac{\hbar}{4 m^2 c^2} \nabla V(\vec{r}) \cdot
(\, \hat{\sigma} \times \vec{p}\, ) \; ,
\end{equation}
where $\hat{\sigma} \equiv \{ \sigma_x , \sigma_y, \sigma_z \}$
denotes Pauli matrices, $V(\vec{r})$ is a
background potential, $\nabla$ stands for the spatial gradient
and $m$ is the electron mass.

Influence of the spin-orbit interaction on the electron transport
properties of two-dimensional mesoscopic systems has been studied
since the early 1980's. At that time it was found that it is
responsible for so-called antilocalization effect \cite{hikami}.
Later the attention has been turned to the effects caused by a Rashba
term \cite{rashba1,rashba2} in two-dimensional \cite{das,doroz,nitta}
and quasi-one-dimensional systems
\cite{entin,aronov,morpurgo,bulgakov,moroz}.
Realistic transport theory for fully quantum coherent systems
including the spin-orbit coupling to the impurities or defects has
not yet been reported. The problem becomes complicated even if
electron motion is restricted to the two-dimensional  space.
In general, the spin-orbit term, Eq.(\ref{soham}), turns the problem
back to three dimensions.

The goal of this paper is to reveal those features of the transport
properties which can be caused by the spin-orbit interaction
induced by a scattering potential. In the
interesting case of a quasi-ballistic regime, which shows chaotic
features, it is very difficult to estimate
deviation from the exact solution caused by any used approximation.
For this reason we have employed simplifying two-component model
with point interaction for which exact solution, including fully
the quantum coherence, can be found. Non-zero spin-orbit coupling
is assumed to be associated with short-range scattering potentials
only. Although
the treatment is far from a realistic transport theory, it might be
useful to understand some mysteries of the recent
experimental observation.

Free electron system is a typical two-component system if
the electron spin is taken into account. If there are no
spin-involved forces, electron states are represented by plane
waves $\exp(i\vec{k}\vec{r})$ with $\vec{k}$  being a wave vector.
Orientation of the electron spin is given by the quantum number
$s_z = \pm${\scriptsize $\frac{1}{2}$} and the electron system
can be splitted into two independent subsystems, each of them
composed of electrons having the same spin orientation.
However, any perturbation of the background potential can cause a
coupling between subsystems due to non-zero spin-orbit term
defined by Eq.(\ref{soham}).

Let us first consider a single scattering potential acting on a
two-dimensional electron gas within a finite region of a radius
$r_{0}$. In accord with standard scattering theory an incoming wave
belonging to one particular subsystem, say of the spin up states,
can be scattered into states belonging to both subsystems.
In the short-range limit, $kr_0<<1$, only $s$-part of the incoming
wave gives non-zero contribution to the scattering process.
For given energy $E=\hbar^2 k^2 / 2 m$, ($k = |\vec{k}|$), the
corresponding solution of the radial Schr\"{o}dinger equation has
two components, $\Psi_{\uparrow}(r)$ and $\Psi_{\downarrow}(r)$.
Outside the scattering region they can be written as follows
\cite{albeverio,exner}
\begin{eqnarray}
\Psi_{\uparrow}(r) \, = \, J_0(kr) \, + \, a(k) H_0^{(1)}(kr) \\
\Psi_{\downarrow}(r) \, = \, b(k) H_0^{(1)}(kr),
\end{eqnarray}
where $J_0(z)$ denotes the cylindrical Bessel function and Hankel
functions $H_0^{(1)}(z)$ represent scattered outgoing waves
that for large arguments have the following asymptotic form
\begin{equation}
H_0^{(1)}(kr \to \infty) \; \sim \;
\frac{1-i}{\sqrt{\pi k }} \;
\frac{e^{ikr}}{\sqrt{r}}  \; .
\end{equation}
Taking into account the time reversal symmetry and assuming that
the system is invariant with respect of the subsystem interchange,
the amplitudes $a(k)$ and $b(k)$ in the
short-range limit have to be of the following general form
\cite{exner}
\begin{eqnarray}
a(k) \, = \, \frac{1 + \frac{2 i}{\pi} \left ( \gamma + \ln
\frac{k}{2} - A \right )}
{\left [ 1 + \frac{2 i}{\pi} \left ( \gamma + \ln
\frac{k}{2} - A \right ) \right]^2
\, + \, \frac{4}{\pi^2} |C|^2} \; ,
\\ \nonumber \\
b(k) \, = \, \frac{2 i}{\pi} C
\left [ 1 + \frac{2 i}{\pi} \left ( \gamma + \ln
\frac{k}{2} - A \right ) \right]^{-1} \, a(k) \; ,
\end{eqnarray}
where $A$ and $C$ are real model parameters. If $C$ is chosen to be
zero, the parameter $A$ represents a strength of the scattering
process within one particular subsystem and for a potential well
of the radius $r_{0}$ it takes the value $A = \ln r_0$. Non-zero
values of the parameter $C$ give rise to a spin-flip
process.

There are two relevant physical quantities
characterizing scattering event: the total scattering
cross-section
%
\begin{equation}
\sigma_0 \, = \, a(k)^2 + b(k)^2
\end{equation}
and the spin-flip probability $t^{\uparrow \downarrow}$
\begin{equation}
t^{\uparrow \downarrow} \, = \, \frac{b(k)^2}{a(k)^2 + b(k)^2} \; .
\end{equation}
Note, that the assumption of the system invariance with respect
of the subsystem interchange leads to the independence of
$\sigma_0$ and $t^{\uparrow \downarrow}$ on the spin orientation
of the incoming electron. This assumption has been used for the
sake of simplicity despite of the fact that it need not to
be satisfied in real systems, e.g. due to a Rashba term.

The scattering problem for a two-dimensional strip with a finite
number of short-range scatterers, as sketched in Fig.~1, can be
solved exactly. The detail analysis in the case of a
one-component system has already been reported \cite{exnerstrip}
and generalization to a two-component system is straightforward.
For simplicity we have assumed that all scatterers are identical,
i.e. they give the same scattering cross-section $\sigma_0$
and spin-flip probability $t^{\uparrow \downarrow}$ if they would
be placed alone  within the two-dimensional space. Scattering
matrix has been obtained numerically for a given configuration
of point scatterers randomly distributed within a strip region
of the length $L$.

\inspicture

Spin-dependent transport properties are determined by partial
transmission coefficients representing transition between left
and right subsystems of asymptotic spin-up or spin-down states.
They are defined as the sum of transmission probabilities
over all relevant modes of asymptotic states.
To simplify the description by excluding the quantum
fluctuations from our consideration we have used configurationally
averaged values of the partial scattering coefficients to define
2$\times$2 transmission and reflection matrices ${\bf T}$ and
${\bf R}$, respectively:
\begin{eqnarray}
{\bf T} \equiv \left(
\begin{array}{cc}
T^{\uparrow \uparrow} &
T^{\uparrow \downarrow} \\
T^{\downarrow \uparrow} &
T^{\downarrow \downarrow}
\end{array} \right) \; , \;
{\bf R} \equiv \left(
\begin{array}{cc}
R^{\uparrow \uparrow} &
R^{\uparrow \downarrow} \\
R^{\downarrow \uparrow} &
R^{\downarrow \downarrow}
\end{array} \right) \; .
\end{eqnarray}
For the considered symmetrical system
$T^{\uparrow \uparrow} \equiv T^{\downarrow \downarrow}$,
$T^{\uparrow \downarrow} \equiv T^{\downarrow \uparrow}$,
$R^{\uparrow \uparrow} \equiv R^{\downarrow \downarrow}$ and
$R^{\uparrow \downarrow} \equiv R^{\downarrow \uparrow}$.

\inspicture

In Fig.~2 and Fig.~3 the dependence of transmission coefficients
on the length $L$ of the scattering region is shown for different
spin-flip probabilities $t^{\uparrow \downarrow}$. The used
energy corresponds to 31 occupied subbands. Concentration of
scatterers (750/$w^2$) and the scattering cross-section
$\sigma_0 = 0.1217$ were held fixed. Probability of an injected
electron to cross scattering region increases with increasing
spin-flip probability as shown in Fig.~2
where the total transmission coefficients
$T = 2 (T^{\uparrow \uparrow} + T^{\uparrow \downarrow})$ are
plotted. This tendency is in agreement with expected
antilocalization effect \cite{hikami}.

\inspicture

The more interesting is the dependence of partial transmission
coefficients. For some values of the spin-flip probability and
lengths $L$, $T^{\uparrow \uparrow}$ becomes less than
$T^{\uparrow \downarrow}$, as can be seen in Fig.~3. It means
that the polarization of the transmitted current has opposite
orientation than the polarization of the injected current.
This surprising result we ascribe to the non-trivial quantum
coherence in two-component systems. Wave interference leading
to the weak localization is one-component effect. It take place
within one particular subsystem only. In the considered case
of a weak spin-flip process ($t^{\uparrow \downarrow} << 1$)
it becomes dominant within the subsystem with
incoming waves while the localization effect within second
subsystem do not need to be for given length $L$ still well
developed.

The above described effect disappears if the scattering
cross-section is substantially enlarged. The localization
becomes dominant effect of the scattering process in the both
subsystems and inequality
$T^{\uparrow \uparrow} > T^{\uparrow \downarrow}$
remains valid for all lengths of the scattering region.

The device conductance of a two-component quantum system is determined by
a matrix of partial transmission coefficients, ${\bf T}_{\rm dev}$,
\begin{eqnarray} \label{conductance}
G = \frac{e^2}{h} (1 \, , 1) \, {\bf T}_{\rm dev}
\left( \begin{array}{c}  1 \\  1 \end{array} \right)
\; ; \;
{\bf T}_{dev} \equiv \left(
\begin{array}{cc}
T_{\rm dev}^{\uparrow \uparrow} &
T_{\rm dev}^{\uparrow \downarrow} \\
T_{\rm dev}^{\downarrow \uparrow} &
T_{\rm dev}^{\downarrow \downarrow}
\end{array} \right) \; ,
\end{eqnarray}
which depends on the properties of ferromagnetic contacts and
their interfaces with the two-dimensional electron gas.
To estimate their effect we have used the idea of
polarization filters \cite{bulgakov}. The source and the
drain are considered to be standard reservoirs and all
spin-dependent effects are modeled by filters placed within
the asymptotic region of ideal leads. If the coherence is
supposed to be completely destroyed at the filter boundaries
the conductance can be expressed as a function of the already
defined  coefficients
$T^{\uparrow \uparrow}$,  $T^{\uparrow \downarrow}$,
$R^{\uparrow \uparrow}$ and $R^{\uparrow \downarrow}$
describing scattering process of the same device without filters.

Experiments on spin-injection into a two-dimensional systems
usually show a large interface resistance. For this reason
non-zero probabilities, $\alpha$ and $\beta$, of spin-up and
spin-down electrons to be reflected by the filter
will be considered. For the sake of the simplicity
we assume that reflected electrons will be equally
distributed between available quantum channels without any
change of their spin orientation.
In this case the filtering effect can be described by 2$\times$2
diagonal matrix
\begin{eqnarray}
{\bf F}_{s,d} \equiv \left(
\begin{array}{cc}
\alpha_{s,d} & 0 \\  0 & \beta_{s,d}
\end{array} \right) \; ,
\end{eqnarray}
where indices $s$ and $d$ represent the
source-filter and the drain-filter, respectively.

For the case of $N$ available quantum channels (subbands)
within each subsystem, the above described model leads to
the following expression for the transmission matrix
${\bf T}_{\rm dev}$ entering the Eq.(\ref{conductance})
\begin{equation}
{\bf T}_{dev} = ({\bf 1} - {\bf F}_{d})
\, {\bf N}{\bf M}_d \, {\bf T} \, {\bf K}_{d,s} \, {\bf N}{\bf M}_s
\, ({\bf 1} - {\bf F}_s)
\end{equation}
where ${\bf N}$ stands for the product of $N$ and unit matrix
${\bf 1}$. The effect of multiple
reflections between filters and the scattering region is
represented by matrices ${\bf M}_{s}$ and ${\bf M}_{d}$
\begin{equation}
{\bf M}_{s} \, = \,
( {\bf N} - {\bf F}_{s} {\bf R})^{-1}
\; ; \;
{\bf M}_{d} \, = \,
( {\bf N} - {\bf R}{\bf F}_{d})^{-1} \; ,
\end{equation}
and
\begin{equation}
{\bf K}_{d,s} \, = \,
[{\bf 1} - {\bf M}_s {\bf F}_{s} {\bf T}
{\bf F}_{d} {\bf M}_d {\bf T}]^{-1}
\; .
\end{equation}

Device conductance, Eq.(\ref{conductance}) depends on
reflection probabilities $\alpha_{s,d}$ and $\beta_{s,d}$
modelling the effect of ferromagnetic contacts.
For the case of the parallel orientation of the contact
magnetization the conductance $G^{\uparrow\uparrow}$ can be obtained
by setting $\alpha_s \equiv \alpha_d$ and
$\beta_s \equiv \beta_d$. To get $G^{\uparrow\downarrow}$ for the
antiparallel contact magnetization $\alpha_s \equiv \beta_d$ and
$\beta_s \equiv \alpha_d$ have to be used. While the conductance
strongly depends on the used values of reflection
probabilities, the sign of the conductance difference
$\Delta G \equiv G^{\uparrow\uparrow} - G^{\uparrow\downarrow}$,
is not affected.

\inspicture

In Fig.~4  the relative conductance change
\begin{equation}
\frac{\Delta G}{G_0} \; \equiv \;
2 \frac{G^{\uparrow\uparrow} - G^{\uparrow\downarrow}}
{G^{\uparrow\uparrow} + G^{\uparrow\downarrow}}
\end{equation}
as a function of the scattering-region length $L$ is shown for
the case of ideal filters, $\alpha_s = 0$ and $\beta_s = 1$.
It corresponds to injection of fully polarized current and
vanishing interface resistance. All other used model parameters
are the same as that for transmission coefficients presented in
Fig.~2 and Fig.~3. The model parameters of the scatterers,
$\sigma_0$ and $t^{\uparrow \downarrow}$,
and their concentration have been chosen to give pronounced
minimum in the dependence $\Delta G / G_0$ on $L$.

To model the device studied by C. M. Hu et al. \cite{hu} with
hundreds occupied subbands it is necessary to
perform calculation for much larger energy and take into account
boundary resistance between ferromagnetic contacts and
two-dimensional electron gas by using non-zero value for
$\alpha_s$. Note that $\alpha_s \sim 0.01$ lowers the values
of the relative conductance change approximately hundred times
and nearly quantitative agreement with the measured data can be
reached as shown in the inset of the Fig.~4.

The presented two-component model does not allow to consider
asymmetry of scattering processes which leads to the spin-Hall
effect \cite{bulgakov}. However, if the polarization
vectors of the contact magnetization lies within the plane of
the electron gas and they are perpendicular to the applied
current, as in the case studied by C. M. Hu et al. \cite{hu},
this effect is suppressed.

The main result of the described model is that in mesoscopic
disordered systems the quantum coherence affected by
spin-flip processes can lead to the higher conductance of
two-terminal devices with antiparallel contact magnetization
than that for parallel configuration as observed in recent
experiments by by C. M. Hu et al. \cite{hu}.
Similar effects can also be expected for other two-component
systems. For example in double-layer systems the defects of
the barrier separating two-dimensional electron gases can act
as their connection points. Instead of a spin-flip process
there will be a real electron transition between electron
layers and in particular cases the conductance measured between
contacts at different layers could be larger than that between
contacts of the same electron layer.

\vspace{-5mm}

\acknowledgements
This work was supported by the Grant Agency of the Czech Republic
under Grant No. 202/98/0085 and by the Academy of Sciences of the
Czech Republic under Grant No. A1048804.
It has been also partially supported by the
"Foundation for Theoretical Physics" in Slemeno, Czech Republic.
The authors acknowledge stimulating discussions with Can Ming Hu.

\end{document}